# Monitoring service quality: methods and solutions to implement a managerial dash-board to improve software development

## A case study in the banking and financial sector

David Luigi FUSCHI
Bridging Consulting Ltd
Reading, UK
david.fuschi@ieee.org

Manuela TVARONAVIČIENĖ
VGTU
Vilnius, Lithuania
manuela.tvaronaviciene@vgtu.lt

Salvatore D'ANTONIO
Universita' degli Studi "Parthenope"
Napoli, Italy
salvatore.dantonio@uniparthenope.it

*Abstract*— **The software used for running and handling the inter-bank network framework provides services with extremely strict uptime (above 99.98 percent) and quality requirements, thus tools to trace and manage changes as well as metrics to measure process quality are essential. Having conducted a two year long campaign of data collection and activity monitoring it has been possible to analyze a huge amount of process data from which many aggregated indicators were derived, selected and evaluated for providing a managerial dash-board to monitor software development.**

## I. INTRODUCTORY CONFIDENTIALITY NOTE

For commercial confidentiality reasons the operative context and location of the experiment cannot be fully disclosed; however, the activity has been conducted in one of the G8 countries within the institutional service provider of the networking and application facilities to the country banking system.

## II. THE CONTEXT

### A. *The institution structure and operation context*

The institution under exam is structured in three main divisions: banking applications, financial markets, and network infrastructure. Each division is responsible for internal software development and maintenance, interaction with subcontractors, service operation, and provision of help-desk services to customers.

Provided services are extremely critical being used to manage the country's financial and banking system. For this reason the provider is subject to strong contractual constraints and commitments (for example a minimum up-time of 99.98%). Based on the criticality of users' request and the status of its internal operations, the scheduling of new releases is decided and planned carefully, and it is essential to evaluate existing relationship among incidents (i.e. customer reported problems), software anomalies (i.e. possible latent, yet not confirmed, errors), and software development so as to make informed decision for the suitability for release of a specific software release.

### B. *Software engineering practice in place and their revision*

Back in 1994 a first software process improvement initiative aimed to monitor and measure products and processes was carried out. This was accomplished within the context of the ESSI project 21244 MIDAS. Activities were mainly focused on Change Management (CM) procedures, policies, and introduction of related tools [2]. Several metrics were regularly collected in relation to software development process and efforts incurred in the development.

Available metrics and tools were assessed through a measurement program based on the Goal-Question-Metrics (GQM) paradigm [3, 1]. A further selection of the most relevant and the introduction of some derived indexes resulted in a significant improvement of the software development practices passing from Capability Maturity Model (CMM) level one to level four.

## III. THE STUDY CARRIED OUT

### A. *Metrics collection object and action field*

The software production and maintenance process relates to a distributed infrastructure providing a full range of services for the management and operation of the national inter-bank network.

The overall system was developed, maintained, and operated by the network division. The dimension of the software under exam amounts to more than five millions Lines of Code (LOC) in C programming language.

The company attention was concentrated on anomalies management and the consequent impact on software maintenance, test and development process across releases.

## B. The board of anomalies

The process to produce high-level indicators and reports evaluating data extracted from collected metrics was initially performed manually. Therefore the first activity performed was the automated extraction of data and report preparation. Anomaly related reports would be the starting point of discussion for Board of Anomalies weekly meeting.

During the Board of Anomalies meetings newly opened and already opened anomalies would be individually discussed. Development team leaders would describe works progresses and provide indication on the availability of a fix (or the expected timeframe to such an event). According to the discussion, test activities would be planned; in particular test-on-demand activities could be requested to investigate anomalies origin. Suggested solution, work around, and information gained would then be spread through the company. Both the result of the discussion and the report produced would also be given to the board of directors along with an historical report depicting the anomalies trend (see Fig. 1 and 2).

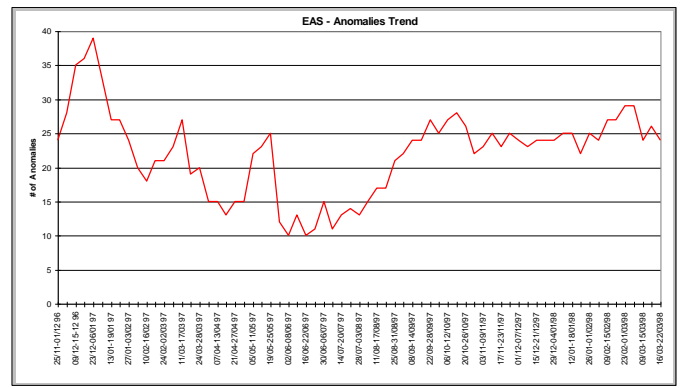

Fig. 1.  Anomaly weekly and overall trend report (per impacted platform)

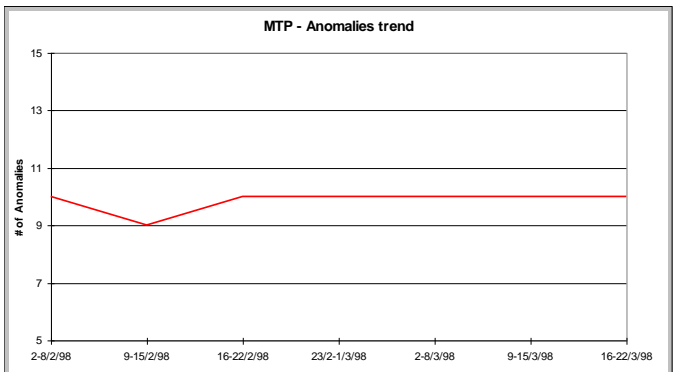

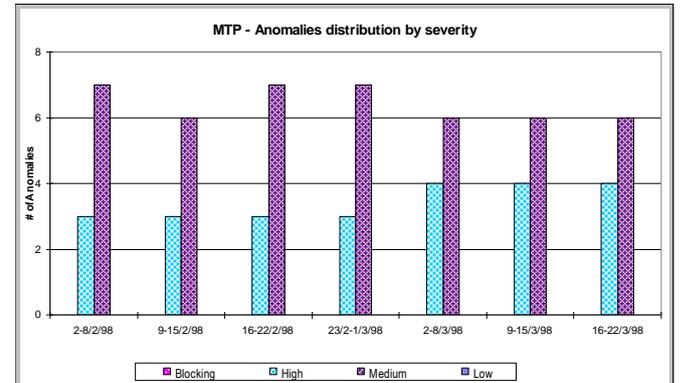

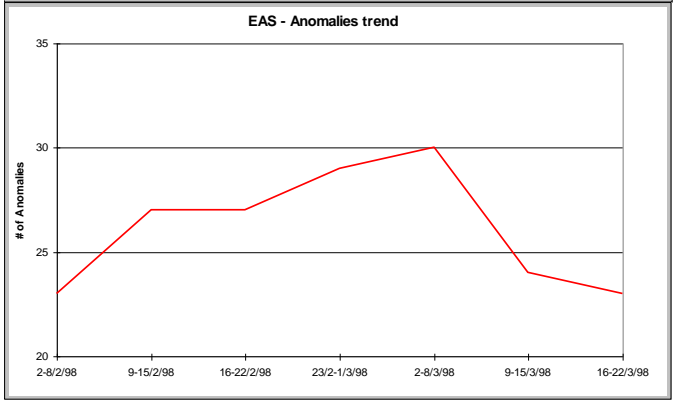

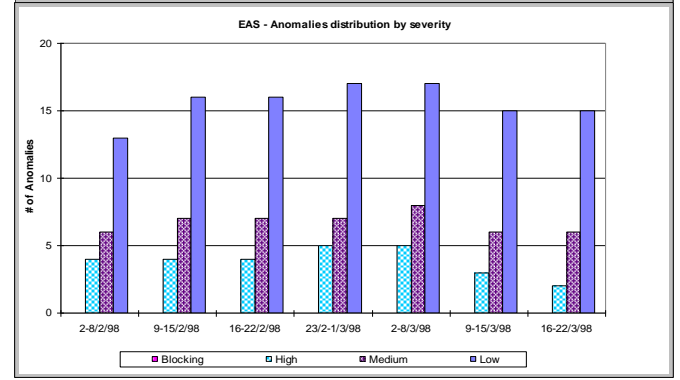

Fig. 2.  Anomaly weekly report (severity details)

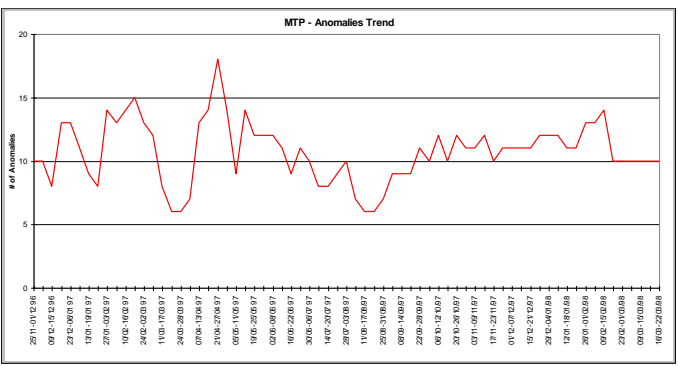

## C. Anomalies report automation

Automation greatly reduced the effort required to produce anomalies report thus enabling its distribution via Intranet. Each project leader / manager could directly access trends and data on-line through a simple dashboard from which to choose what to see and how (overall counts, details, trends, etc.).

According to [11] the number of residual errors decreases with the testing coverage. Therefore when considering a series of subsequent releases of the same software it is reasonable to expect that an initial peak of anomalies should be found in correspondence with a new re-lease (or a fix) due to introduced changes and enhancements, followed by a constant decrease in anomalies number as testing proceeds. In other words, it is reasonable to expect an overall monotonically decreasing trend towards a lower bound due to the fact that it will be impossible to totally avoid errors [11]. Such behavior is shown in Fig 3.

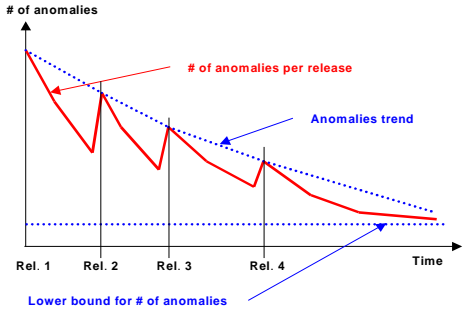

Fig. 3. Theorical behaviour of anomalies trend

On the other hand if the actual trend, as observed from field measures, is substantially different from the theoretical one, an accurate investigation should be carried out.

## IV. THE METRICS REVISION

### A. Indicators selection criteria

While many metrics can be collected, not all of them will be of use. Furthermore, different metrics are of interest to different kind of people. For example, project managers and technicians will be interested in KLOC (1000 Lines Of Code), number of faults, MTBF (Mean Time Between Failures), etc.; line managers will be interested in Productivity, Quality, Cost, On-time-delivery, etc.; finally, top managers will be interested in aggregated metrics useful to support decision taking. According to Basili [3] metrics can be derived by adopting the model presented in Fig.4 hereafter.

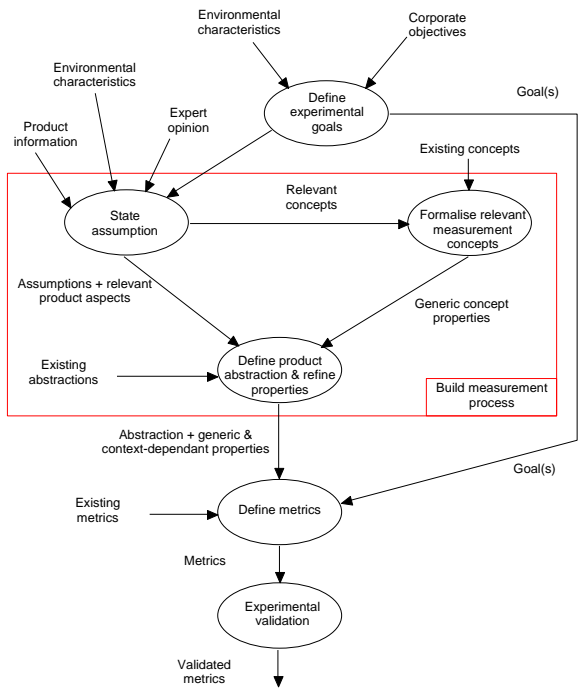

Fig. 4. Basili Model

The application of Basili model [3] proved very effective, although it was necessary to add a feed-back loop in the process so as to refine both the assumptions and the set of metrics to be used following the comments and requests placed by managers engaged in the process. This allowed achieving the best possible decision making support environment.

### B. Software releases observed

The experiment has been conducted processing metrics collected for the following system components:

- "MTP" - the front-end access point to the transport network (usually resident on a UNIX ® machine).
- "EAS" - the actual application front-end access point to the network (an application usually resident on customer's mainframe environment and interfacing with one or more MTP).

To ensure the management could base its decisions on a correct evaluation of provided indicators, only data related to software releases used in actual operation have been considered (i.e. releases that were not used in production environment have not been considered).

Indexes have been classified in major categories according to the measured object or derivation metric. Examined classes can be summarized as follows:

TABLE I. ANALYZED INDEXES CATEGORIES

| Product quality | Process quality | Size & Productivity |
|---|---|---|
| • MTBF<br>• Total failures<br>• Failure rate<br>• Integration failure rate<br>• Quality index<br>• Availability | • MTTF<br>• Efficiency degree<br>• Test quality index<br>• Maintenance test time for KLOC | • KLOC<br>• KLOCC<br>• Product change rate |

## V. ADOPTED INDICATORS EXPLANATION AND USAGE

Most Key Performance Indicators (KPI) adopted regarding product development life cycle quality assurance have been based on theory (i.e. as per normative literature) [5, 8, 9, 10].

However, there are problems that are not described (or only partially addressed) in literature, therefore some indicators were adapted (or even derived anew) based on observation and practical experience. Indicators selected are reported hereafter.

### A. Mean Time To Failures

It shows the mean elapsed time between two consecutive faults [5, 9]. It is computed as the ratio between total life time (*TL*) and total number of failures during life (*LA*). An increasing trend denotes an improvement in the reliability of the software.

$$MTTF = TL / LA \qquad (1)$$

*B. Mean Time To Repair*

It gives a synthetic view of the quality of software repair process [5]. A low value implies either that anomalies are of minor relevance or that the repair process is efficient.

Here *TNA* is the total number of anomalies found for the examined release at the time being.

A high value denotes that anomalies solution is difficult. This is usually the case when a product is mature and therefore remaining uncovered bugs are difficult to find or solve.

$$MTTR = [\Sigma_{i=1,TNF} (t_{AN\ closure} - t_{AN\ opening})] / TNA \quad (2)$$

*C. Mean Time Between Failures*

It gives a synthetic view of the quality of software [5]. In fact if *MTTF* is high and *MTTR* is low, the overall process is running smoothly.

It can be used for predicting fault occurrence only if derived from a long series of historical data. Anyhow as these are mean values, derived indication should be compared with anomalies distribution, failure rate, effort and elapsed time per process phase.

$$MTBF = MTTF + MTTR \quad (3)$$

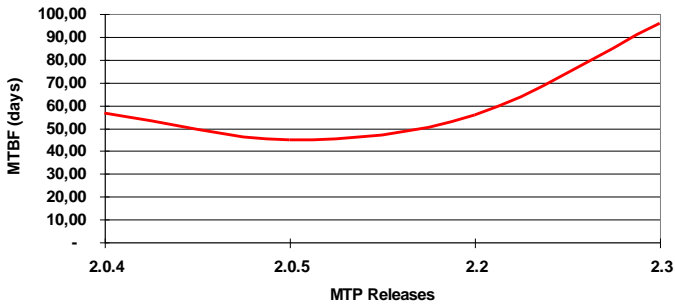

*D. Total Failures*

It shows the normalized ratio between failures during life and the total amount of changed lines in code [9]. A decreasing trend denotes an improvement in software, only if there is an overall decrease in total number of failures during life (*LA*) as well as an increase in *MTTF*.

In the specific case *KLCC* stands for *KLOC* changed in code that we have computed as the sum of newly inserted lines with changed and deleted ones.

$$TF / KLOC = LA / KLCC \quad (4)$$

*E. Failure Rate*

Provides an intuitive indication of how the system is running as it is computed as the ratio between total number of failures during life (*LA*) and total life time (*TL*).

If it is decreasing the process is running correctly and the product quality and reliability is in-creasing [9, 10].

$$FR = LA / TL \quad (5)$$

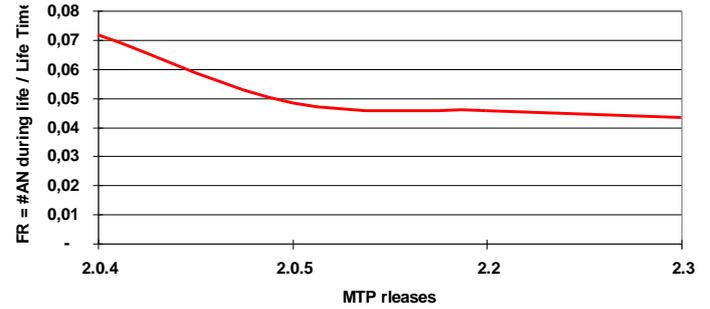

*F. Quality index*

This is a very simple indication of product quality if plotted for subsequent releases [5, 10]. It is simple to compute and can give a clear indication of how the product development process is running. If everything is correct it will present a monotonically de-creasing trend.

$$Quality = Defect / KLOC \quad (6)$$

*G. Availability*

This index provides an average indication of the product availability between failures [5]. It is computed taking into account both *MTTF* and *MTTR*.

If the development process was behaving correctly this indicator should decrease in time (it grows linearly with increasing *MTTF* and decreasing *MTTR*, while it increases in a logarithmic form with increasing *MTTF* and increasing *MTTR*).

$$AV\ \% = (MTTF) / (MTTF + MTTR)\ \% \quad (7)$$

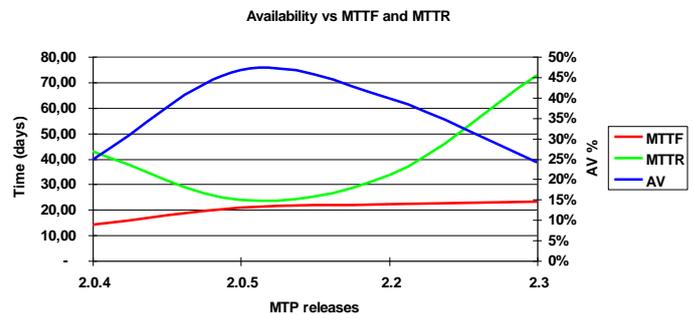

*H. Efficiency Degree*

It gives an indication of the efficiency of the product as it compares Product Change Rate (*PCR*) to total product life time (*TL*).

The smaller this ratio the more efficient have been product revision / maintenance process [10].

$$ED = PCR / TL \qquad (8)$$

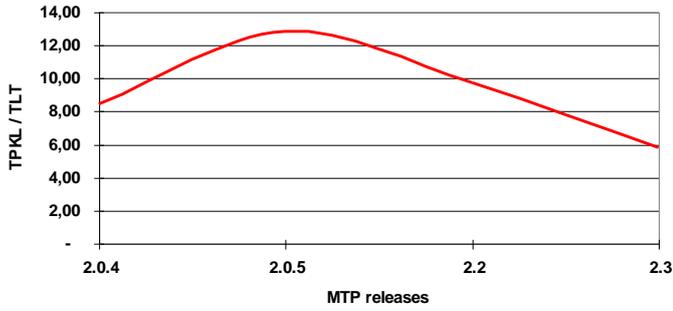

### I. Integration Failure Rate

It gives an indication on the quality of the product under integration [9]. It is computed as the ratio of number of failures during test (*TA*) to test time ($T_{Test}$). Retaining constant test time, a growth in its value denotes a decreased quality of product. While a decreasing value denotes an increased quality of the product.

$$IFR = TA / T_{Test} \qquad (9)$$

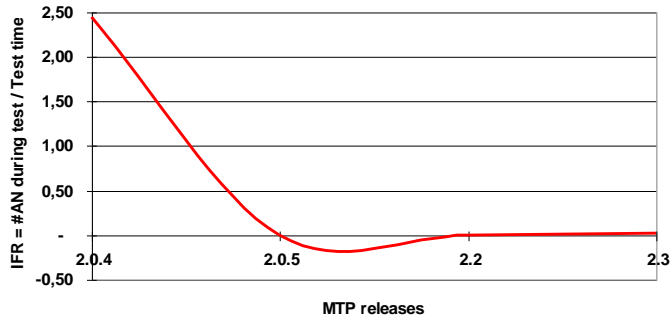

### J. Test Quality Index

Provides an indication of the test process quality as it is derived from the ratio of number of failures during life (*LA*) with number of failures during test (*TA*). Its decrease means that the quality of the overall test process is increasing.

$$TQI = LA / TA \qquad (10)$$

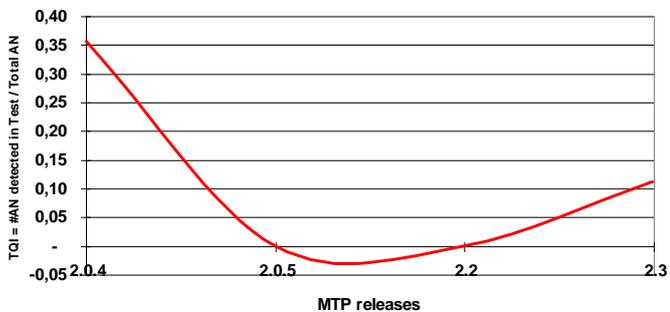

### K. Maintenance Test Time for KLOC

It shows how test time (*TTEST*) for *KLOC* changed in code (*KLCC*) varies [9]. An increase in its value means a heavier test effort or a more difficult test phase. A decrease suggests a reduced test activity or a very large increase in changes. In any case such value has to be compared with integration failure rate (*IFR*) and failure rate (*FR*) in order to gain proper information on the quality of the test process.

$$MTT / KLOC = TTEST / KLCC \qquad (11)$$

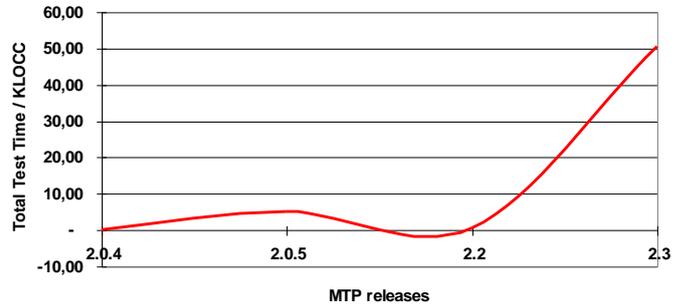

### L. Product Change Rate

It gives an indication on the rate of change in the product [8]. To compute this index is required to perform the ratio between *KLOC* changed in code (*KLCC*) and total product *KLOC* (*TPKL*). *KLCC* has been considered as the sum of new, changed and deleted lines.

$$PCR = KLCC / TPKL \qquad (12)$$

### M. KLOC versus Function Points (FP)

Many indexes have been derived from *KLOC* data even if most of them could have also been derived from *FP* data. *KLOC* was preferred as *FP*s are more difficult to be evaluated. According to [5], *FP* can be roughly derived starting from *KLOC* using the following rule:

$$FP = LOC_{application} / [(LOC/FP) * BAF] \qquad (13)$$

where *LOC/FP* is an average value dependent on the programming language used for development (in this case C programming language in which case the value is 120), while BAF is a backfiring adjustment factor related to complexity, here assumed to be 1.30 for complex applications [5].

Other indicators proposed in [5], such as the Productivity Index[1] or the Documentation Index[2] were also investigated but considered as not suitable for the purposes of this study.

Other useful indications can be derived by examining anomalies distribution, efforts and phase times over a series of releases.

---

[1] *PI = KLOC / Effort*
[2] *DI = man pages / KLOC*

As far as anomalies are concerned the most interesting information can be gained by looking at:

- anomalies distribution per release (Fig.5, 6),
- severity distribution,
- detection environment (internal test, external test or production – Fig.6, 7).

Examining anomalies distribution implies assessing how many new anomalies have been found for each release, how many have been inherited (anomalies already found but not jet solved when the new code was released) and how many have been solved. Release specific data can be seen as indicators of product behavior, while average values as process related indexes.

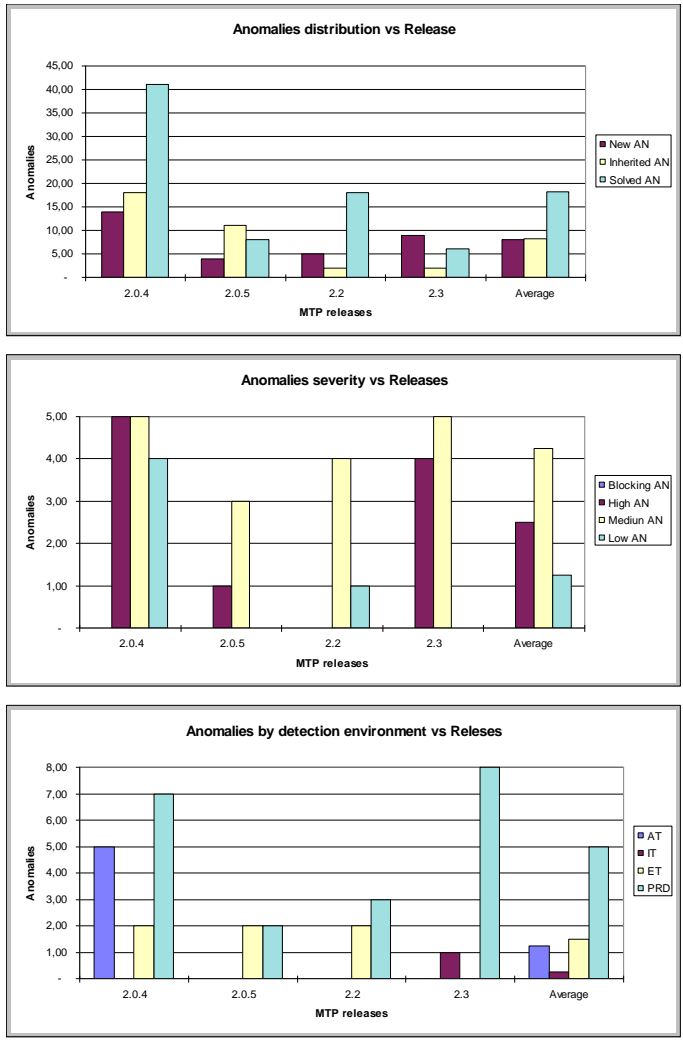

Fig. 5. Anomalies related information

Severity distribution is particularly important as it gives a more in depth view on the kind of problems experienced. In the examined case it was found that on average were encountered more medium severity anomalies than high ones, moreover it was found that no blocking anomalies were detected over a series of consecutive releases.

Anomalies detection environment data can be used to evaluate development processes when compared with test effort and time. This also gives a direct indication on how difficult it is to replicate field condition while testing.

Interesting data on the overall development process evolution could be gained by observing time distribution per phase as it allows monitoring maintenance impact in terms of development and test time over product life time.

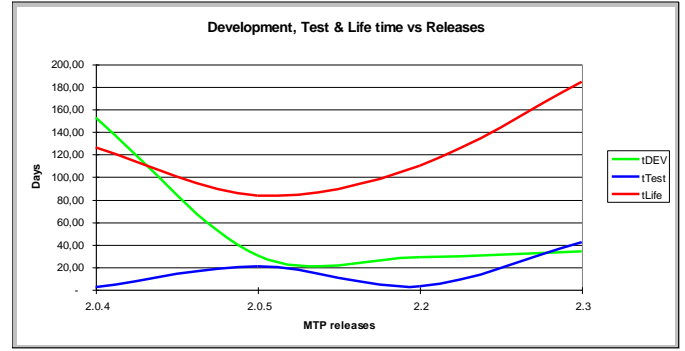

Fig. 6. Development, test and life-time versus anomalies

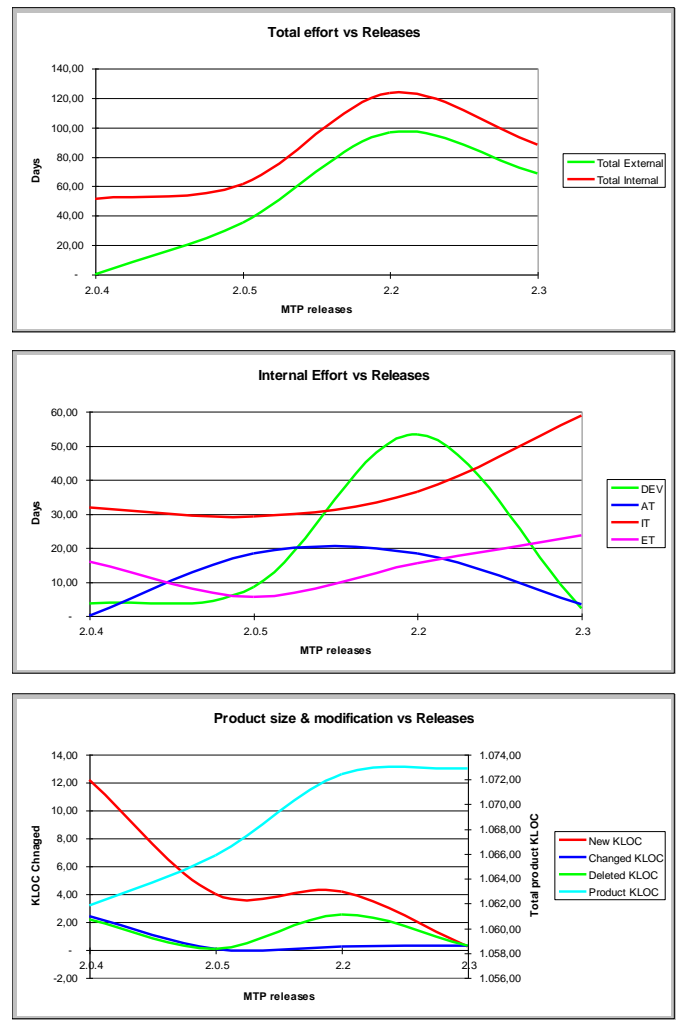

Fig. 7. Effort and product change data versus release

Other relevant data on the development process could be gained by examining both effort distribution per phase and overall effort distribution. In particular, effort data should always be used when examining process phases duration and anomalies trends. This is due to the fact that maintenance process is made of recursive development steps followed by test and validation ones. If the elapsed time of a specific phase appears to have been squeezed, it will be necessary to verify if in the same phase efforts have been significantly increased in order to avoid output has been compromised. At the same time a final assessment of the overall effectiveness of adopted countermeasures can only be gained by assessing the quality of the phase outputs as measurable with the previously described indicators.

## VI. ON-LINE SCOREBOARD DATA PROVISION

At first the attention focus was on index derivation and therefore tools like Excel® and SPSS® were used to process data extracted from metrics and effort database (implemented in Oracle®). Data extraction was achieved using Oracle ProC© and SQL tools. Most pre-processing computation was carried out in Oracle ProC© as it was allowing to select and process data on line (see Fig. 8). As a first step derived data were exported to plain field delimited ASCII files (with semicolons ";" as separator) and then imported in Excel® and SPSS® where further analysis and graphic generation could be easily accomplished.

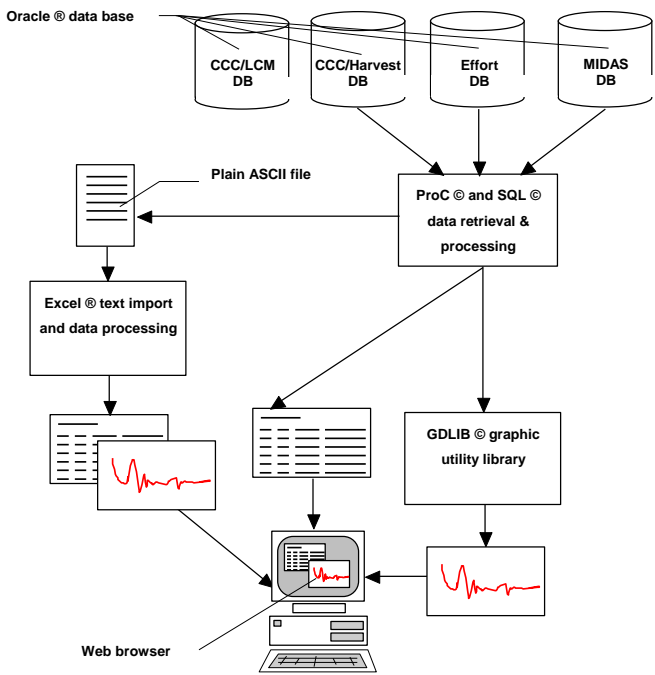

Fig. 8. Adopted data processing process

The second step was a further processing one in which retrieved data is indexed and graphically plotted using GDLIB© graphic utilities so as to enable plotted data usage in the designed web-based scoreboard system. Anomalies report, metrics and indexes suite could be tailored on demand thanks to the aforementioned process, and made available to a web server via a specifically available CGI interface. Users could then access data online via a simple browser.

Given the nature of available data and indexes it was possible to compute the average, the standard deviation, correlation and/or regression index. Such statistical operation could be request for all values or a specific subset. For example it would be possible to evaluate the correlation between MTTR and effort or between PCR, MTTR and Quality, thus allowing experienced users to get a quantitative evaluation of hypothesis arisen from metrics inspection and qualitative evaluation.

## VII. CONCLUSIONS

The conducted experiment has enabled to derive a working tool with a stepwise approach. It has proved that adopted quality control policies and data collection practices need to be constantly refined. It has also showed that statistical analysis has to be carried out on collected data in order to gain even more info from the past. Furthermore, this work has allowed a significant decrease in the effort required to prepare management meetings and a significant in-crease in:

- timely available information;
- level of confidence in decision;
- level of control over processes.

Based on the results of this work and the gained experience, it is possible to formulate the following recommendation:

- any software development company should base its software process on a sound and reliable CM environment;
- different level of detail / aggregation should be available as different are the needs of professional involved in day by day operation;
- GQM paradigm should be used in defining indicators and what to track;
- metrics should be collected but they must be identified according to company's need;
- indicators should be easily derivable form collected metrics;
- no unnecessary overhead should be introduced in the monitored process.

In conclusion it is possible to argue that the original idea to attempt deriving from well-established and consolidated quality assurance mechanism and metrics in use for machinery maintenance metrics and indicators valid also for software development and maintenance, has proved viable and beneficial, to the extent that the adoption of the proposed monitoring of the software development and maintenance process (that was part of a quality assurance and testing automation effort) allowed passing from a ratio of 20% of anomalies identified in internal testing and 80% identified in external testing and/or production to a 80% identified in internal testing and 20% in external testing and/or production with a substantial increase in the management and customers satisfaction.